\documentclass[12pt]{article}
\usepackage{amsfonts}
\usepackage{amsmath}
\usepackage{amssymb}
\usepackage{graphicx}
\usepackage[all, knot]{xy}
\usepackage{tikz}
\usepackage{array}

\usepackage{ulem}

\usepackage[utf8]{inputenc}
\usepackage{epstopdf}
\usepackage[footnotesize]{caption}
\usepackage{amsthm}
\usepackage{enumitem}
\usepackage{mathrsfs}

\usepackage[margin=3cm]{geometry}

\def \be {\begin{equation}}
\def \ee {\end{equation}}
\def \bea {\begin{eqnarray}}
\def \eea {\end{eqnarray}}
\def \nn {\nonumber}

\def \rr {\raise.35ex\hbox{\small $\prime$}\kern-.17em{\mbox{\large $\imath$}}}

\def \dels {\partial\kern-.6em /\kern.1em}
\def \As {{A\kern-.5em / \kern.5em}}
\def \Ds {D\kern-.7em / \kern.5em}

\def \ks {k\kern-.5em /}
\def \ls {l\kern-.5em /}

\newcommand{\ci}[1]{}
\newcommand{\ba}{\begin{eqnarray}}
\newcommand{\ea}{\end{eqnarray}}
\newcommand{\bal}{\begin{align}}
\newcommand{\eal}{\end{align}}
\newcommand{\bay}[1]{\left(\begin{array}{#1}}
\newcommand{\eay}{\end{array}\right)}

\newcommand{\ket}[1]{|{#1}\rangle}

\def\xD{{\Delta}}

\def\xG{{\Gamma}}

\newcommand{\hide}[1]{}

\newlist{axioms}{enumerate}{2}
\setlist[axioms,1]{label=\textbf{A\arabic{axiomsi}.}, ref=A\arabic{axiomsi}}
\setlist[axioms,2]{label=\textbf{A\arabic{axiomsi}\rlap{\myEnumCounter{axiomsii}}.},%
                   ref=A\arabic{axiomsi}\myEnumCounter{axiomsii},%
                   align=parleft,%
                   leftmargin=0em,%
                   itemsep=1.4ex,%
                   before={\stepcounter{axiomsi}}}

  \usetikzlibrary{decorations.markings}

\begin{document}
\begin{titlepage}
%\begin{flushright}
%NORDITA 2019-102 %\\
%October,  2019
%\end{flushright}

\begin{center}

\textbf{\LARGE
dS/CFT Correspondence from\\ 
a Defect Operator
\vskip.3cm
}
\vskip .5in
{\large
Xing Huang$^{a, b}$ \footnote{e-mail address: xingavatar@gmail.com} and
Chen-Te Ma$^{c}$ \footnote{e-mail address: yefgst@gmail.com}, 
\\
\vskip 1mm
}
{\sl
$^a$
Institute of Modern Physics, Northwest University, Xi'an 710069, China.
\\
$^b$
NSFC-SPTP Peng Huanwu Center for Fundamental Theory, Xi'an 710127, China. 
\\
$^c$
Department of Physics, Great Bay University, Dongguan, Guangdong 52300, China. 
}
\\
\vskip 1mm
\vspace{40pt}
\end{center}

%\newpage
\begin{abstract}
\noindent
We perform a Wick rotation and analytic continuation from global AdS$_{d+1}$ to static dS$_{d+1}$, yielding CFT$_d$ generators with a nonstandard adjoint action tied to dS bulk coordinates. 
To reproduce the real-scalar two-point function, we introduce a global defect operator that twists the inner product. 
We further show that $PT$ symmetry is spontaneously broken in CFT$_2$ vacua with a central charge having an imaginary part. 
Finally, we derive integral identities for bulk and defect correlators, providing a unified framework for computing CFT$_d$ observables in the presence of global and local defects.
\end{abstract}
\end{titlepage}

\section{Introduction}
\label{sec:1}
\noindent
The anti-de Sitter/conformal field theory (AdS/CFT) correspondence provides a non-perturbative definition of quantum gravity in asymptotically AdS spacetimes \cite{Maldacena:1997re,Gubser:1998bc,Witten:1998qj}.
A central question is whether a similar framework can be formulated for dS space.
One approach is analytic continuation from AdS to dS \cite{Strominger:2001pn,Anninos:2011ui}; however, such continuation alone does not fully determine the quantum structure of the dual theory.
\\

\noindent
Because a conformal field theory possesses scale invariance, it is often better defined than the corresponding bulk gravitational theory.
This motivates an alternative perspective in which gravity is emergent rather than fundamental.
In this approach, a gravitational description arises only in the large-central-charge limit of the CFT.
The proper quantum gravity regime should be explored—or even defined—entirely within the CFT framework \cite{Balasubramanian:1998sn,Banks:1998dd}.
Within this program, an important objective is to understand the mechanism by which a bulk field emerges and to determine the specific conditions a CFT must satisfy to exhibit such emergent gravitational behavior.
A central tool in this investigation is the Hamilton-Kabat-Lifschytz-Lowe (HKLL) bulk reconstruction procedure \cite{Hamilton:2006az,Czech:2015qta,Huang:2019wzc,Huang:2020cye}.
This method reconstructs bulk AdS fields from non-local combinations of boundary CFT operators, thereby providing a concrete map between bulk dynamics and boundary observables and enabling the study of gravitational phenomena directly from the CFT perspective.
\\

\noindent
To understand our universe from the perspective of quantum gravity, it is essential to study CFT associated with a de Sitter (dS) background.
One approach is to {\it analytically continue} the AdS curvature radius \cite{Strominger:2001pn},
\bea
L \rightarrow iL,
\eea
and correspondingly continue the positive central charge of CFT$_d$ \cite{Brown:1986nw,Cotler:2018zff,Huang:2019nfm,Huang:2020tjl},
\bea
c \propto \frac{L^{d-1}}{G_N} \ \longrightarrow\ i^{d-1}\frac{L^{d-1}}{G_N},
\eea
where $G_N$ is the physical gravitation constant \cite{Strominger:2001pn}.
This continuation implies, in particular, that the central charge of CFT$_2$ has an imaginary part \cite{Strominger:2001pn}.
Indeed, the {\it adjoint} operation of the CFT$_2$ generators, when represented in dS$_3$ bulk coordinates, exhibits {\it exotic} properties \cite{Doi:2024nty}.
\\

\noindent
To reproduce the correct real scalar two-point function \cite{Bousso:2001mw,Ishibashi:1988kg,Nakayama:2015mva,Bhowmick:2019nso}, it becomes necessary to {\it modify} the inner-product structure of the bulk local state space \cite{Ishibashi:1988kg}.
This requires introducing a biorthogonal basis of eigenstates \cite{Doi:2024nty}.
The need for such a modification already follows from the dS bulk isometry group, which signals both the alteration of the inner product and the {\it loss} of conventional hermiticity.
By incorporating parity ($P$) and time-reversal ($T$) operations as part of the adjoint structure for bulk local states, one is naturally led to define a {\it global} defect operator $G_D$ acting on the vector space $\mathcal{H}$, characterized by the involutive property
\bea
G_D^2 = 1.
\eea
Consequently, the dS/CFT correspondence can be formulated in the language of {\it non-Hermitian} quantum theory \cite{Bender:1998ke,Bender:2002vv,Mostafazadeh:2001jk,Mostafazadeh:2001nr,Mostafazadeh:2002id,Mostafazadeh:2002maq} and in terms of {\it defect} operators.
\\

\noindent
We have another kind of defect operator, the {\it local} defect operator, which resides on a $p$-dimensional submanifold, exhibits a remarkably universal structure governing how broken internal symmetries imprint themselves on local defect observables.
Ward-identity analyses show that CFT correlators obey integral relations \cite{Drukker:2022pxk} connecting {\it higher}-point correlators to {\it lower}-point data, enabling powerful nonperturbative constraints across diverse models \cite{Belton:2025ief} and informing perturbative studies of O($N$) line defects \cite{Belton:2025ief,Girault:2025kzt}.
Complementary geometric paradigms elucidate these relationships as reflections of a defect conformal manifold, within which the Zamolodchikov metric and curvature are intricately encoded in the two- and four-point functions of precisely marginal defect operators \cite{Drukker:2022pxk,Barrat:2021tpn,Cavaglia:2022qpg}.
Analytic bootstrap techniques further demonstrate that such integral identities sharpen operator unmixing and constrain anomalous dimensions in local defect spectra \cite{Belton:2025hbu}.
\\

\noindent
The presence of a defect explicitly breaks a global symmetry to a subgroup, resulting in a modified Ward identity.
The integral identity provides a general and systematic framework for classifying these modifications \cite{Belton:2025ief}.
We use the term anomaly in the standard sense: a symmetry of the classical theory that is not preserved at the quantum level due to regularization effects.
In defect CFTs, the non-conservation of the bulk current is localized on the defect and governed by a specific operator whose transformation properties uniquely determine the defect contribution to the Ward identity.
This defect-localized term captures a universal obstruction to extending the bulk CFT symmetry across the defect.
Crucially, the anomaly terms, such as the Weyl anomaly, can be addressed generically only through the integral identity, which provides a computationally advantageous tool unmatched by alternative methods \cite{Belton:2025ief}.
Consequently, the integral identity serves as a powerful method for computing higher-point connected correlators without loss of generality \cite{Belton:2025ief}.
This analysis unifies geometric and algebraic perspectives on defect conformal manifolds, yielding explicit expressions for anomaly coefficients and new constraints on defect CFT data.
\\

\noindent
Most previous AdS$\rightarrow$dS continuation focus on the bulk partition functions \cite{Strominger:2001pn}, wavefunction coefficients \cite{Maldacena:2011mk}, cosmological correlators \cite{McFadden:2011kk,Bzowski:2011ab,Bzowski:2012ih}, and the higher spin dualities \cite{Anninos:2011ui}. 
The present work instead focuses on Hilbert-space structure and the adjoint operation.
We show that the analytical continuation changes the adjoint structure of conformal generators.
Because the ordinary Hermitian conjugation fails to reproduce the dS scalar Green function, a modified inner product is required.
A global defect operator can represent this modification.
The resulting theory naturally fits into a non-Hermitian/$PT$-symmetric framework.
Unlike previous continuation programs, the essential new ingredient is not merely the continuation of correlators but the continuation of the Hilbert-space metric itself.
In higher-spin dS/CFT, the dual CFT is non-unitary and possesses unusual Hilbert-space properties.
Our analysis suggests that such non-unitarity possibly be understood more generally as arising from a modified inner product encoded by a global defect operator.
Although we do not construct the higher-spin dual explicitly, the $PT$-defect framework developed here may provide a concrete realization of the Hilbert-space structure that has remained somewhat mysterious in higher-spin dS/CFT.
In the cosmological holography program, analytic continuation relates AdS correlation functions to inflationary observables and late-time wavefunctions \cite{McFadden:2011kk,Anninos:2014lwa,Sleight:2019mgd}.
Our goal differs conceptually. 
In contrast, we study the realization of the conformal algebra itself and investigate the resulting adjoint operation. 
The necessity of introducing a $PT$ defect therefore appears at the level of Hilbert-space structure rather than at the level of correlation functions alone.

\subsection{Summary of Results}
\noindent
In this paper, we investigate the symmetry perspective of the dS/CFT correspondence obtained through Wick rotation and analytic continuation.
Although considerations of symmetry alone are generally inadequate to ascertain a comprehensive quantum theory—owing to quantum corrections or anomalies—we assert that Wick rotation and analytic continuation can nevertheless be meaningfully employed within a circumscribed set of structures.
Our analysis focuses specifically on the metric, the CFT generators associated with the bulk isometry group, and physical observables defined through expectation values.
The action, by contrast, is subtle under such continuations because quantum effects can modify it in nontrivial ways.
Nevertheless, the restricted continuation we adopt suffices to extract structural information about de Sitter physics directly from the CFT side, and our derivations remain valid for CFTs in arbitrary dimensions.
While analytic continuation from AdS to dS has been extensively studied, our work identifies an additional necessary structure: the modification of the inner product in the dual CFT.
We show that this modification can be naturally formulated as a global defect operator that twists the Hilbert space structure.
This leads to a non-Hermitian realization of the conformal algebra and provides a consistent framework for reproducing de Sitter bulk correlators.
Our main results are summarized as follows (as in Fig. \ref{ds_cft_summary_trimmed}):
\begin{figure}[t]
\centering
\includegraphics[width=0.72\textwidth]{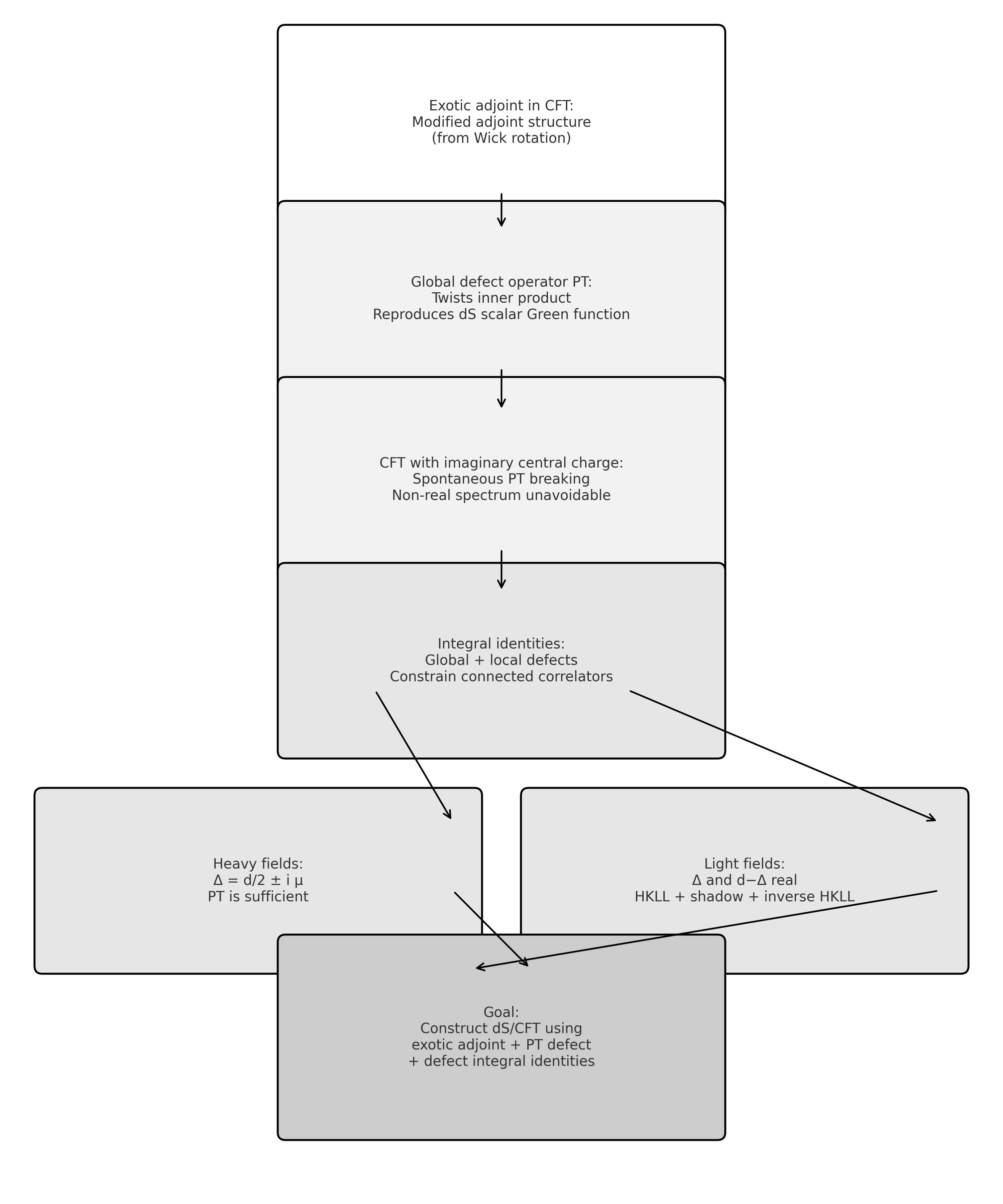}
\caption{Schematic summary of the dS/CFT construction in this work: from Wick-rotated AdS to static dS, exotic adjoint and global $PT$ defect, spontaneous $PT$ symmetry-breaking, and integral identities with global and local defects constraining CFT bulk and defect correlators.}
\label{ds_cft_summary_trimmed}
\end{figure}
\begin{itemize}
\item We first derive the static patch of dS$_{d+1}$ from Lorentzian global AdS$_{d+1}$ via Wick rotation and analytic continuation, and then identify the corresponding CFT$_d$ generators.
We show that the adjoint operation in this setting differs from the familiar structure appearing in the AdS/CFT correspondence.
\item We study bulk local states in the static patch of dS$_{d+1}$, noting that the same construction applies to the global patch as well.
By computing the two-point Green’s function of a real scalar field, we demonstrate that reproducing this result from the CFT$_d$ side requires introducing a global defect—or equivalently, a $PT$ operator—to twist the CFT inner product.
The introduction of a global defect is not optional but required: without modifying the inner product, the CFT cannot reproduce the correct de Sitter Green’s functions.
\item We show that $PT$ symmetry must be spontaneously broken for any vacuum state in a CFT$_2$ with a central charge having an imaginary part.
This implies the unavoidable presence of non-real operator dimensions in the dS/CFT correspondence.
\item We demonstrate how such a CFT remains computable under integral identities in the presence of both global and local defects.
This clarifies how global defects are incorporated beyond two-point functions and how conformal symmetry continues to impose constraints on the defect conformal group.
\end{itemize}

\noindent
The structure of the paper is as follows.
In Sec.~\ref{sec:2}, we derive the static dS$_{d+1}$ geometry via Wick rotation and describe the associated CFT$_d$ generators and adjoint operation.
In Sec.~\ref{sec:3}, we construct bulk local states for both the static and global patches of dS.
In Sec.~\ref{sec:4}, we introduce the global defect operator $PT$ and show how it twists the CFT$_d$ inner product to reproduce the dS$_{d+1}$ scalar two-point function.
In Sec.~\ref{sec:5}, we prove that spontaneous $PT$ symmetry breaking is inevitable in any CFT$_2$ with a central charge having an imaginary part.
In Sec.~\ref{sec:6}, we present the integral identity and demonstrate its application in the presence of global and local defects.
We conclude in Sec.~\ref{sec:7} with a summary and discussion of our results.

%The main results of our work are presented in Fig. \ref{summary.png}.
%\begin{figure}
%\begin{center}
%\includegraphics[width=1.\textwidth]{summary.png}
%\end{center}
%\caption{The result's summarization and implication. }
%\label{summary.png}
%\end{figure}

\section{Static dS from Global AdS}
\label{sec:2}
\noindent
In this section, we first review some basics like the static dS metric for later convenience. 
The main result is to show that the conformal generators for dS/CFT in general dimensions obey a non-standard adjoint operation. 
Let us begin introducing the embedding coordinates $X^A$,
\bea
\eta^{AB}X^AX^B=-X^2_{-1}-X_0^2+\sum_{j=1}^dX_j^2=-L^2,
\eea
where the metric is
\bea
\eta^{AB}=\mathrm{diag}(-1, -1, 1, 1, \cdots, 1),
\eea
and the $L$ is the radius of curvature.
We label the embedding coordinates' indices as $A=-1, 0, 1, \cdots, d$.
The global coordinates ($\tau, \rho, \Omega_{d-1}$) are:
\bea
X_{-1}=L\cosh(\rho)\cos(\tau); \
X_0=L\cosh(\rho)\sin(\tau); \ \cdots; \
X_j=L\sinh(\rho)\Omega_j,
\eea
where
\bea
\sum_{j=1}^d\Omega_j^2=1.
\eea
The range of $\tau$ is from $-\infty$ to $\infty$, the range of $\rho$ is $\lbrack 0, \infty)$, and $\Omega_j\in S^{d-1}$.
The AdS$_{d+1}$ metric in the global coordinates is
\bea
ds^2=L^2\big(-\cosh^2(\rho)d\tau^2+d\rho^2+\sinh^2(\rho)d\Omega_{d-1}^2\big).
\eea
We can consider $\rho\rightarrow\infty$ to approach the AdS$_{d+1}$ conformal boundary $\mathbb{R}\times S^{d-1}$ (a cylinder).
The isometry group is SO($d$, 2), and the boundary conformal group is also SO($d$, 2).
For $d=2$, the AdS$_3$ has the isometry group SO(2, 2)$\cong$SL(2, $\mathbb{R}$)$_L$$\otimes$SL(2, $\mathbb{R}$)$_R$.
This represents the overarching symmetry of AdS$_3$, about the global conformal group of CFT$_2$.
Three elements can generate each SL(2, R) factor:
\bea
(L_0, L_1, L_{-1}); \ (\tilde{L}_0, \tilde{L}_1, \tilde{L}_{-1}).
\eea
The conformal algebra so(2, 2) can also be written in terms of the so(2, 1)$\bigoplus$so(2,1) generators, each satisfying
\bea
\lbrack L_0, L_{\pm 1}\rbrack=\mp L_{\pm 1}; \
\lbrack L_1, L_{-1}\rbrack=2L_0.
\eea
We write the CFT$_2$ generators when using the AdS bulk coordinates representation:
\bea
&&
L_0=\frac{i}{2}(\partial_{\tau}+\partial_{\phi}), \
L_{\pm 1}=\frac{i}{2}e^{\pm i(\tau+\phi)}
\bigg(\frac{\sinh\rho}{\cosh\rho}\partial_{\tau}
+\frac{\cosh\rho}{\sinh\rho}\partial_{\phi}\mp i\partial_{\rho}\bigg);
\nn\\
&&
\tilde{L}_0=\frac{i}{2}(\partial_{\tau}-\partial_{\phi}), \
\tilde{L}_{\pm 1}=\frac{i}{2}e^{\pm i(\tau-\phi)}
\bigg(\frac{\sinh\rho}{\cosh\rho}\partial_{\tau}
-\frac{\cosh\rho}{\sinh\rho}\partial_{\phi}\mp i\partial_{\rho}\bigg).
\eea
The algebra satisfies:
\bea
\lbrack L_m, L_n\rbrack=(m-n)L_{m+n}; \ \lbrack \tilde{L}_m, \tilde{L}_n\rbrack=(m-n)\tilde{L}_{m+n}.
\eea
The Hermitian conjugate in the algebra shows:
\bea
(L_n)^{\dagger}=L_{-n}; \ (\tilde{L}_n)^{\dagger}=\tilde{L}_{-n}.
\eea
\\

\noindent
We can do the Wick rotation and analytical continuation from the global AdS$_{d+1}$ to the static dS$_{d+1}$,
\bea
\tau\rightarrow it; \ \rho\rightarrow i\theta; \ L^2\rightarrow-L^2,
\eea
to obtain the static dS$_{d+1}$ metric
\bea
ds^2=L^2\big(-\cos^2(\theta)dt^2+d\theta^2+\sin^2(\theta)d\Omega_{d-1}^2\big),
\eea
where
\bea
0<\theta<\pi; \ -\infty<t<\infty.
\eea
In the dS/CFT correspondence, the dual conformal field theory is posited on the sphere at the asymptotic future and past infinities, represented as $t=\pm\infty$.
We also apply the Wick rotation and analytical continuation to the CFT$_2$ generators:
\bea
&&
L_0=\frac{1}{2}(\partial_{t}+i\partial_{\phi}), \
L_{\pm 1}=\frac{i}{2}e^{\pm i(-t+i\phi)}
\bigg(\frac{\sin\theta}{\cos\theta}\partial_{t}
-i\frac{\cos\theta}{\sin\theta}\partial_{\phi}\mp i\partial_{\rho}\bigg);
\nn\\
&&
\tilde{L}_0=\frac{1}{2}(\partial_{t}-i\partial_{\phi}), \
\tilde{L}_{\pm 1}=\frac{i}{2}e^{\pm i(-t-i\phi)}
\bigg(\frac{\sin\theta}{\cos\theta}\partial_{t}
+i\frac{\cos\theta}{\sin\theta}\partial_{\phi}\mp i\partial_{\rho}\bigg).
\eea
The algebra satisfies:
\bea
\lbrack L_m, L_n\rbrack=(m-n)L_{m+n}; \ \lbrack \tilde{L}_m, \tilde{L}_n\rbrack=(m-n)\tilde{L}_{m+n}.
\eea
The Hermitian conjugate in the algebra shows the non-conventional property \cite{Doi:2024nty}:
\bea
&&
(L_0)^{\dagger}=-\tilde{L}_{0}, \ (L_{\pm 1})^{\dagger}=\tilde{L}_{\pm 1};
\nn\\
&&
(\tilde{L}_0)^{\dagger}=-L_{0}; \ (\tilde{L}_{\pm 1})^{\dagger}=L_{\pm 1}.
\eea 
\\

\noindent
To consider a general dimension, we have the CFT$_d$ generators: $P_{\mu}$ are the translation generators; $M_{\mu\nu}$ are the Lorentz generators; $D$ generates the scaling transformations (dilatons); $K_{\mu}$ generate the special conformal transformations.
In general dimensions, a real scalar field with a given mass $m$, which satisfies
\bea
m^2L^2>\frac{d^2}{4}, 
\eea 
we have two conformal dimensions for the dS$_{d+1}$ case
\bea
\Delta_{\pm}=\frac{d}{2}\pm\sqrt{\frac{d^2}{4}-m^2L^2}\equiv\frac{d}{2}\pm i\mu,
\eea
which can be obtained via Wick rotation and analytical continuation from the AdS$_{d+1}$ case.
The conformal dimensions are the eigenvalues of the $D$ operator.
The adjoint operation acting on these generators for the dS$_{d+1}$ bulk coordinate representation is
\bea
D^{\dagger}=-D; \
K_{\mu}^{\dagger}=K_{\mu}; \
P_{\mu}^{\dagger}=P_{\mu}.
\eea
When the conformal generators act on the primary states $\ket{\Delta_\pm}$, which is the lowest-weight state, the adjoint operation does not have the conventional form:
\bea
&&
\langle\hat{\Delta}_{\pm}|D^{\dagger}=\langle\Delta_{\pm}|\hat{\Delta}_{\mp}; \
D^{\dagger}|\hat{\Delta}_{\pm}\rangle=\Delta_{\mp}|\hat{\Delta}_{\pm}\rangle; \
0=\langle\hat{\Delta}_{\pm}|K_{\mu}^{\dagger}=P^{\dagger}_{\mu}|\hat{\Delta}_{\pm}\rangle;
\nn\\
&&
\langle\hat{\Delta}_{\pm}|D=-\langle\hat{\Delta}_{\pm}|\Delta_{\mp}; \
D|\hat{\Delta}_{\pm}\rangle=-\Delta_{\mp}|\hat{\Delta}_{\pm}\rangle; \
0=\langle\hat{\Delta}_{\pm}|K_{\mu}=P_{\mu}|\hat{\Delta}_{\pm}\rangle,
\eea
where the normalziation is:
\bea
\langle\Delta_j|\Delta_k\rangle=\langle\hat{\Delta}_j|\hat{\Delta}_k\rangle=\delta_{jk}; \
\langle\Delta_j|\hat{\Delta}_k\rangle=\langle\hat{\Delta}_j|\Delta_k\rangle=0.
\eea
We define the adjoint bra state as:
\bea
(|\Delta_{\pm}\rangle)^{\dagger}\equiv\langle\hat{\Delta}_{\pm}|; \
(\langle\Delta_{\pm}|)^{\dagger}=|\hat{\Delta}_{\pm}\rangle.
\eea

\section{Bulk Local State}
\label{sec:3}
\noindent
In this section, we construct bulk local states in dS$_{d+1}$, which are CFT states corresponding to local scalar field operators in the bulk. 
We also demonstrate that antipodal symmetry exchanges the two conformal dimensions $\Delta_{\pm}$ related to the same mass. The bulk local state for any dS$_{d+1}$ is \cite{Nakayama:2015mva}
\bea
\label{bklocal}
|\Psi_{\Delta}\rangle= \sum_{n=0}^{\infty} (-1)^n C_n (P^2)^n | \Delta \rangle,
\eea
where $|\Delta\rangle$ is a conformal primary state, and
\bea
P^2 = \sum_{a=0}^{d-1} P_a P_a; \
C_n = \prod_{k=1}^n \frac{1}{4k \Delta + 4 k^2 -2kd}.
\eea
It satisfies \cite{Nakayama:2015mva}:
\bea
M_{\mu\nu}|\Psi_\Delta\rangle=0; \ (P_\mu + K_\mu)|\Psi_\Delta\rangle =  0.
\label{blsv}
\eea
The bulk local state is equivalent to using an AdS$_{d+1}$ bulk scalar field acting on the vacuum state \cite{Nakayama:2015mva}.
We can use the Wick rotation and the analytical continuation to transition to the static dS$_{d+1}$, and find that the bulk local state satisfies the same relation \eqref{blsv}.
Indeed, we can perform the coordinate transformation from the static coordinate system to the global coordinate system.
Therefore, the bulk local states are the same between the static and global coordinates \cite{Doi:2024nty}.
Hence, our results using the bulk local state will be the same between the static and global dS$_{d+1}$ \cite{Doi:2024nty}.
\\

\noindent
We can apply $\exp(Dt){\cal R}\exp(i\theta J)$, where
\bea
J\equiv\frac{1}{2}(K_0-P_0),
\eea
to a bulk local state to take any point to anywhere on a sphere, $S^{d-1}$.
Since we have
\bea
e^{i\pi J}De^{-i\pi J}=-D,
\eea
we obtain
\bea
\langle\hat{\Delta}_{\pm}|e^{i\pi J}D=\Delta_{\mp}\langle\hat{\Delta}_{\pm}|e^{i\pi J},
\eea
whcih shows that $\langle\hat{\Delta}_{\pm}|e^{i\pi J}$ is proportional to a conjugate transpose of the primary state $\langle\Delta_{\mp}|$,
\bea
\langle\hat{\Delta}_{\pm}|e^{i\pi J}=\nu_{\mp}\langle\Delta_{\mp}|.
\eea
Taking the conjugate transpose, we obtain
\bea
e^{-i\pi J}|\Delta_{\pm}\rangle=\nu_{\mp}^*|\hat{\Delta}_{\mp}\rangle.
\eea
Hence, we obtain that
\bea
\nu_{\pm}\langle\Delta_{\pm}|e^{-2\pi J}|\Delta_{\pm}\rangle=\nu_{\mp}^*.
\eea
When we set all spacetime variables to zero, except for $\theta$, we can find that $\exp(i\theta J)$ acting on the bulk local state gives the shift $\theta\rightarrow \theta+\pi$, which implies that the operation maps a dS bulk point to the antipodal point.
The antipodal and dS bulk points in a sphere are the endpoints of a diameter that a straight line segment connects them through the sphere's center.
Because $\exp(i\pi J) |\Psi_{\xD}\rangle$ remains a bulk local state, the state is the eigenstate of the operator, and we then use
$\exp(D t) {\cal R}$ to take a dS bulk point to anywhere, these operations realize the following equality
\bea
|\Psi_{\Delta_{\pm}}(t, \theta+\pi, \Omega)\rangle=\lambda_{\pm}|\Psi_{\Delta_{\pm}}(t, \theta, \Omega)\rangle.
\eea
\\

\noindent
The sum in the definition of the bulk local state \eqref{bklocal} can be carried out and gives \cite{Nakayama:2015mva}
\bea
\label{bulklocald}  |\Psi_\Delta\rangle = \xG\bigg(\Delta-\frac{d}{2}+1\bigg) \left(\frac{\sqrt{P^2}}{2}\right)^{\frac{d}{2}-\Delta} J_{\Delta-\frac{d}{2}} (\sqrt{P^2})|\xD \rangle\,,
\eea
which agrees with the known HKLL form \cite{Hamilton:2006az,Czech:2015qta,Huang:2019wzc,Huang:2020cye} (in momentum space representation) of a bulk scalar field acting on a vacuum state.
It can be further shown from the dictionary of the AdS/CFT correspondence \cite{Witten:1998qj} that this particular form is enough to get the bulk-to-boundary propagator, and hence the two-point function of a real scalar 
\bea
G(x, y)=\frac{\Gamma(\Delta)}{2\pi^{\frac{d}{2}}\Gamma\big(\Delta-\frac{d}{2}+1\big)}
e^{-\Delta D_{\mathrm{AdS}}(x, y)} {}_2F{}_1\bigg(\Delta, \frac{d}{2}, \Delta+1-\frac{d}{2}; e^{-2D_{\mathrm{AdS}}(x, y)}\bigg),
\eea
where the Gamma function is
\bea
\Gamma(z)\equiv\int_0^{\infty}dt\ t^{z-1}e^{-t}, \ \mathrm{Re}(z)>0,
\eea
and the hypergeometric function is
\bea
{}_2F_1(a, b, c; z)\equiv\sum_{n=0}^{\infty}\frac{(a)_n(b)_n}{(c)_n}\frac{z^n}{n!}, \ |z|<1.
\eea
The Pochhammer symbol is defined as
\bea
(a)_n\equiv\left\{\begin{array}{ll}
1, & \mbox{$n=0$}. \\
a(a+1)\cdots (a+n-1), & \mbox{$n>0$}.
\end{array} \right.
\eea
The $D_{\mathrm{AdS}}(x, y)$ is the geodesic line for the AdS$_{d+1}$ background between two points, $x$ and $y$.
When $|z|\ge 1$, we define the hypergeometric function by the analytical continuation directly.
We can use the Wick rotation and the analytical continuation to obtain the following Green's function
\bea
G_{\Delta_{\pm}}(x, y)=\frac{\Gamma(\Delta_{\pm})}{2\pi^{\frac{d}{2}}\Gamma\big(\Delta_{\pm}-\frac{d}{2}+1\big)}
e^{-i\Delta_{\pm} D_{\mathrm{dS}}(x, y)} {}_2F{}_1\bigg(\Delta_{\pm}, \frac{d}{2}, \Delta_{\pm}+1-\frac{d}{2}; e^{-2iD_{\mathrm{dS}}(x, y)}\bigg),
\label{gf1}
\nn\\
\eea
where $D_{\mathrm{dS}}(x, y)$ is the geodesic line for the dS$_{d+1}$ background connecting two points $x$ and $y$, and it connectes to $D_{\mathrm{AdS}}(x, y)$ as
\bea
D_{\mathrm{AdS}}(x, y)=iD_{\mathrm{dS}}(x, y).
\eea
According to Eq. \eqref{gf1}, we can read
\bea
\frac{1}{\lambda_{\pm}}=(-1)^{\frac{d}{2}}e^{\pm\pi\mu}
\eea
from:
\bea
\langle\Psi_{\Delta_{\pm}}(0, \theta+\pi, 0)|\Psi_{\Delta_{\pm}}(0, 0, 0)\rangle
&=&\langle\Psi_{\Delta_{\pm}}(0, \theta, 0)|  e^{-i\pi J} |\Psi_{\Delta_{\pm}}(0, 0, 0)\rangle
\nn\\
&=&(-1)^{\frac{d}{2}}e^{\pm\pi\mu}\langle\Psi_{\Delta_{\pm}}(0, \theta, 0)|\Psi_{\Delta_{\pm}}(0, 0, 0)\rangle.
\eea
Hence, we can obtain
\bea
\langle\Delta_{\pm}|e^{-2\pi J}|\Delta_{\pm}\rangle=(-1)^de^{\pm 2\pi\mu},
\eea
and obrain
\bea
\nu_{\pm}=(-1)^{\frac{d}{2}+1}e^{\mp\pi\mu}.
\eea
\\

\noindent
We can take the conjugate transpose of the bulk local state to get:
\bea
\langle\hat{\Psi}_{\Delta_{\pm}}|&=&
\bigg\langle\hat{\Delta}_{\pm}\bigg|\sum_{n=0}^{\infty}(-1)^nC_n(\Delta_{\mp})(P^2)^ne^{-i\theta J}{\cal R}^{-1}e^{-Dt}
\nn\\
&=&
\bigg\langle\hat{\Delta}_{\pm}\bigg|\sum_{n=0}^{\infty}(-1)^nC_n(\Delta_{\mp})(P^2)^ne^{i\pi J}e^{-i\pi J}
e^{-i\theta J}{\cal R}^{-1}e^{-Dt}
\nn\\
&=&\bigg\langle \hat{\Delta}_{\pm}\bigg|e^{i\pi J}\sum_{n=0}^{\infty}(-1)^nC_n(\Delta_{\mp})(K^2)^ne^{-i(\theta+\pi) J}{\cal R}^{-1}e^{-Dt}
\nn\\
&=&\nu_{\mp}\bigg\langle \Delta_{\pm}\bigg|\sum_{n=0}^{\infty}(-1)^nC_n(\Delta_{\mp})(K^2)^ne^{-i(\theta+\pi) J}{\cal R}^{-1}e^{-Dt}
\nn\\
&=&\nu_{\mp}\langle\Psi_{\Delta_{\mp}}(t, \theta+\pi, \Omega)|.
\eea
Hence, we show that the inner product of the non-vanishing bulk local states is
\bea
&&
\langle\hat{\Psi}_{\Delta_{\mp}}(x)|\Psi_{\Delta_{\pm}}(y)\rangle
\nn\\
&=&\frac{\nu_{\pm}\Gamma(\Delta_{\pm})}{2\pi^{\frac{d}{2}}\Gamma\big(\Delta_{\pm}-\frac{d}{2}+1\big)}
e^{-i\Delta_{\pm} D_{\mathrm{dS}}(x_A, y)} {}_2F{}_1\bigg(\Delta_{\pm}, \frac{d}{2}, \Delta_{\pm}+1-\frac{d}{2}; e^{-2iD_{\mathrm{dS}}(x_A, y)}\bigg),
\eea
where $x_A$ is the antipodal point, with respect to the dS bulk point $x$, and the other inner product of the bulk local states vanishes
\bea
\langle\hat{\Psi}_{\Delta_{\pm}}(x)|\Psi_{\Delta_{\pm}}(y)\rangle=0.
\eea
The bulk local states $|\Psi_{\Delta_{\pm}}\rangle$ both correspond to the same mass, and we want to make a linear combination to obtain another Green's function with the dual to the dS$_{d+1}$ free scalar field that we explain in the next section.

\section{Green's Function}
\label{sec:4}
\noindent
The principal result of this section is to show that the correct Wightman function can be reproduced from the inner product between bulk local states only after introducing a $PT$-twist. 
The Green's function (Wightman function) for a real scalar field in a Bunch-Davies vacuum state takes the following form \cite{Bousso:2001mw}
\bea
G_E(x, y)=\frac{\Gamma(\Delta_+)\Gamma(\Delta_-)}{(4\pi)^{\frac{d+1}{2}}\Gamma\big(\frac{d+1}{2}\big)}
{}_2F_1\bigg\lbrack\Delta_+, \Delta_-, \frac{d+1}{2}; \cos^2\bigg(\frac{D_{\mathrm{dS}}(x, y)}{2}\bigg)\bigg\rbrack.
\eea
We can use the following identities:
\bea
\cos^2\bigg(\frac{D_{\mathrm{dS}}(x_A, y)}{2}\bigg)
=\sin^2\bigg(\frac{D_{\mathrm{dS}}(x, y)}{2}\bigg)
=1-\cos^2\bigg(\frac{D_{\mathrm{dS}}(x, y)}{2}\bigg),
\eea
where
\bea
D_{\mathrm{dS}}(x_A, y)=\pi-D_{\mathrm{dS}}(x, y),
\eea
\bea
{}_2F_1\bigg(a, b, \frac{a+b+1}{2}; 1-z\bigg)
={}_2F_1\bigg(\frac{a}{2}, \frac{b}{2}, \frac{a+b+1}{2}; 4z(1-z)\bigg)
\eea
with
\bea
z=\cos^2\bigg(\frac{D_{\mathrm{dS}}(x, y)}{2}\bigg)
\eea
to obtain
\bea
G_E(x_A, y)=\frac{\Gamma(\Delta_+)\Gamma(\Delta_-)}{(4\pi)^{\frac{d+1}{2}}\Gamma\big(\frac{d+1}{2}\big)}
{}_2F_1\bigg(\frac{\Delta_+}{2}, \frac{\Delta_-}{2}, \frac{d+1}{2}; \sin^2\big(D_{\mathrm{dS}}(x, y)\big)\bigg).
\eea
We then apply the identity
\bea
{}_2F_1\bigg(a, b, 2b; z\bigg)=(1-z)^{-\frac{a}{2}}{}_2F_1\bigg(\frac{a}{2}, b-\frac{a}{2}, b+\frac{1}{2}; \frac{z^2}{4(z-1)}\bigg)
\eea
to rewrite $G_E(x_A, y)$ as
\bea
G_E(x_A, y)=\frac{\Gamma(\Delta_+)\Gamma(\Delta_-)}{(4\pi)^{\frac{d+1}{2}}\Gamma\big(\frac{d+1}{2}\big)}
e^{i\Delta_+ D_{\mathrm{dS}}}
{}_2F_1\bigg(\Delta_+, \frac{d}{2}, d; 1-e^{2iD_{\mathrm{dS}}}\bigg).
\eea
The $G_E(x_A, y)$ can also be rewritten as the symmetric form by exchaning $\Delta_+$ and $\Delta_-$,
\bea
&&
G_E(x_A, y)
\nn\\
&=&\frac{\Gamma(\Delta_-)\Gamma\big(\frac{d}{2}-\Delta_-)}{4(\pi)^{\frac{d}{2}+1}}
e^{i\Delta_-D_{\mathrm{dS}}(x, y)}
{}_2F_1\bigg(\Delta_-, \frac{d}{2}, -\frac{d}{2}+\Delta_-+1; e^{2i D_{\mathrm{dS}}(x, y)}\bigg)
\nn\\
&&+\frac{\Gamma(\Delta_+)\Gamma\big(\frac{d}{2}-\Delta_+)}{4(\pi)^{\frac{d}{2}+1}}
e^{i\Delta_+D_{\mathrm{dS}}(x, y)}
{}_2F_1\bigg(\Delta_+, \frac{d}{2}, -\frac{d}{2}+\Delta_++1; e^{2i D_{\mathrm{dS}}(x, y)}\bigg)
\nn\\
\eea
by using the following formula
\bea
&&
{}_2F_1(a, b, c; z)
\nn\\
&=&\frac{(1-z)^{-a-b+c}\Gamma(c)\Gamma(a+b-c)}{\Gamma(a)\Gamma(b)}
{}_2F_1(c-a, c-b; -a-b+c+1; 1-z)
\nn\\
&&
+\frac{\Gamma(c)\Gamma(-a-b+c)}{\Gamma(c-a)\Gamma(c-b)}
{}_2F_1(a, b, a+b-c+1; 1-z).
\eea
From this symmetric form, we can use the linear combination of the primaries $\Delta_{\pm}$ to write the dual wavefunction
\bea
|\Psi_E(x)\rangle=\frac{1}{\sqrt{2\sinh(\pi\mu)}}\bigg(\frac{1}{\sqrt{i}}|\Psi_{\Delta_+}(x)\rangle+\sqrt{i}|\Psi_{\Delta_-}(x_A)\rangle\bigg),
\eea
and then $G_E(x, y)$ can be obtained by replacing $x$ with $x_A$ in the $D_{\mathrm{dS}}(x, y)$ from $G_E(x_A, y)$ and can be written as an inner product:
\bea
G_E(x, y)
=\langle\Psi_E(x)|\Psi_E(y)\rangle
=\frac{i}{2\sinh(\pi\mu)}
\bigg(\langle\hat{\Psi}_{\Delta_+}(x)|\Psi_{\Delta_-}(y_A)\rangle
-\langle\hat{\Psi}_{\Delta_-}(x_A)|\Psi_{\Delta_+}(y)\rangle\bigg),
\nn\\
\eea
where
\bea
\langle\Psi_E(x)|=\frac{1}{\sqrt{2\sinh(\pi\mu)}}\bigg(\sqrt{i}\langle\hat{\Psi}_{\Delta_+}(x)|
+\frac{1}{\sqrt{i}}\langle\hat{\Psi}_{\Delta_-}(x_A)|\bigg).
\eea
\\

\noindent
Finally, let us discuss the implications of the result from the perspective of the inner product.
We observe that the bra state can be obtained from the $PT$ transformation:
\bea
\nu_{-}\langle 0|PT(PT)^{-1}\Phi_{\Delta_+}(t, \theta, \Omega)PT
=\nu_{-}\langle 0|_{PT}\Phi_{\Delta_-}(t, \theta+\pi, \Omega)
=\langle\hat{\Psi}_{\Delta_+}(x)|,
\eea
where
\bea
\langle 0|_{PT}\equiv\langle 0|PT.
\eea
The $\nu_{\mp}$ can be thought of as the linear combination of the coefficients for the states
\bea
\langle\Psi_E(x)|=\frac{1}{\sqrt{2\sinh(\pi\mu)}}\bigg(\sqrt{i}\nu_-\langle\hat{\tilde{\Psi}}_{\Delta_+}(x)|
+\frac{\nu_+}{\sqrt{i}}\langle\hat{\tilde{\Psi}}_{\Delta_-}(x_A)|\bigg),
\eea
where
\bea
\nu_{\mp}\langle\hat{\tilde{\Psi}}_{\Delta_{\pm}}|=\langle\hat{\Psi}_{\Delta_{\pm}}|.
\eea
We can also use the conventional $PT$ inner product to define $\langle\hat{\tilde{\Psi}}_{\Delta_{\pm}}|$,
\bea
\langle\hat{\tilde{\Psi}}_{\Delta_{\pm}}(x_A)|=\langle\Psi_{\Delta_{\pm}}(x)|PT.
\eea
Hence, it is equivalent to twisting the inner product space by the $PT$ transformation.
The $PT$ transformation maps a dS bulk point to an antipodal point and also exchanges the conformal dimensions, $\Delta_+\leftrightarrow\Delta_-$.
We then observe that the $PT$ operation acting on the $|\Psi_E\rangle$ is invariant due to that
\bea
PT\bigg(\frac{1}{\sqrt{i}}|\Psi_{\Delta_+}(x)\rangle\bigg)\longleftrightarrow \sqrt{i}|\Psi_{\Delta_-}(x_A)\rangle; \
PT\big(\sqrt{i}|\Psi_{\Delta_-}(x_A)\rangle\big)\longleftrightarrow\frac{1}{\sqrt{i}}|\Psi_{\Delta_+}(x)\rangle.
\eea
Hence, the linear combination makes the $PT$ invariant state, so that we can make the conjugation transpose for obtaining the bra state, but the complex conjugate of $|\Psi_{\Delta_{\pm}}\rangle$ corresponds to the different conformal dimensions, so that we need to twist the inner product by $PT$ operator, which is one class of the global defect for spinless particles
\bea
(PT)^2=1.
\eea
Because the defect operator corresponds to a symmetry of a theory, the global defect acting on a vacuum state is proportional to the vacuum state if we do not have spontaneous symmetry breaking.
However, the sphere has antipodal symmetry, so we expect the vacuum state to have degeneracies, and we should observe that each vacuum state's conjugate spectrum corresponds to another vacuum state's spectrum.
In other words, we expect that spontaneous symmetry breaking is generic for the CFT's vacuum states.
We demonstrate the proof for the vacuum states in any CFT$_2$.

\section{Spontaneous $PT$ Symmetry Breaking}
\label{sec:5}
\noindent 
The main theorem of this section is that a CFT$_2$ with complex central charge cannot possess a $PT$-invariant vacuum. 
Since the Virasoro generators are
\bea
L_n=\frac{1}{2\pi i}\oint_{|z|=1} dz\ z^{n+1}T(z),
\eea
the parity transformation reverses the contour, and the time reversal transformation conjugates the coefficient:
\bea
(PT)L_n(PT)^{-1}&=&\frac{1}{2\pi i}\oint_{|\bar{z}|=1}d\bar{z}\ \bar{z}^{n+1}(PT)T(z)(PT)^{-1}
=\frac{1}{2\pi i}\oint_{|\bar{z}|=1}d\bar{z}\ \bar{z}^{n+1}T^{\dagger}(\bar{z})
\nn\\
&=&
\frac{1}{2\pi i}\oint_{|z|=1}dz\ z^{n+1}T^{\dagger}(z)
=L_n^{\dagger}.
\eea
The Virasoro algebra is
\bea
\lbrack L_m, L_n\rbrack=(m-n)L_{m+n}+\frac{c}{12}(m^3-m)\delta_{m+n, 0},
\eea
where $c$ is a central charge with an imaginary part.
After we apply the $PT$ transformation on the left and right sides, the Virasoro algebra becomes
\bea
\lbrack L_m^{\dagger}, L_n^{\dagger}\rbrack=(m-n)L_{m+n}^{\dagger}+\frac{c^*}{12}(m^3-m)\delta_{m+n, 0}.
\eea
The physical vacuum must satisfy
\bea
L_n|0\rangle=0, \ n\ge0.
\eea
If the vacuum state is also a $PT$ eigenstate
\bea
PT|0\rangle=\lambda|0\rangle,
\eea
we apply the $PT$ operator to obtain:
\bea
0=PTL_{n\ge0}|0\rangle=(PT)L_{n\ge0}(PT)^{-1}(PT)|0\rangle=L_{n\ge0}^{\dagger}\lambda|0\rangle.
\eea
Hence, we get the condition
\bea
L_{n\ge0}^{\dagger}|0\rangle=0.
\eea
Because the algebra of $L_n$ is the Virasoro algebra with a central charge $c$, and the algebra of $L_n^{\dagger}$ is the Virasoro algebra with central charge $c^*$, and the vacuum cannot simultaneously satisfy the highest weight conditions for $c$ and $c^*$, the vacuum state cannot be the $PT$ invariant or in the $PT$ symmetric phase for the central charge with an imaginary part.
The precise mathematical statement is that we use the following relation
\bea
\langle 0|\lbrack L_{-m}-L^{\dagger}_{-m}, L_{m\ge0}-L^{\dagger}_{m\ge0}\rbrack|0\rangle=\frac{c-c^*}{12}m(m^2-1)=0
\eea
to prove that the $PT$ invariant vacuum state only exists when the central charge is a real number ($c=c^*$).
Hence, the vacuum state must transform into a different state under the $PT$ transformation, implying that the vacuum state is not an eigenstate of the $PT$ operator.
This proof implies that the non-real spectrum is unavoidable in the dS/CFT correspondecne.

\section{Integral Identity of Connected Correlators}
\label{sec:6}
\noindent
In this section, we demonstrate that the global defect operator can be incorporated into the integral-identity formalism of defect CFTs. 
We introduce the CFT bulk operators and global and local defect operators simultaneously for the most general situations.
Since the introduction of a global defect operator in CFT$_d$ should depend on what bulk theory we consider, we generically use $G_D$ as the global defect for twisting an inner product.
For the free dS bulk scalar case, the theory respects the $PT$ symmetry, and we introduce the global defect operator, $PT$.
However, even for the free theory with charges, we expect the inner product to change upon introducing the charge-conjugation operator.
Hence, we consider the most generic approach here.
We introduce the flat defect, the generating function, and the integral identity in the presence of the local defect \cite{Belton:2025ief}.
We then discuss how to introduce global defect operators to have the dS$_{d+1}$ dual from the CFT$_d$ and demonstrate how the conformal symmetry is still applicable to determine the connected correlators for the conformal defect group case.

\subsection{Flat Defect}
\noindent
We consider a $d$-dimensional Euclidean CFT with a global internal symmetry group $G$ and insert a flat $p$-dimensional conformal defect $W$ (bare defect operator) along a flat submanifold
\bea
\mathbb{R}^p\subset\mathbb{R}^d
\eea
and coordinates
\bea
x=(\tau^a, x_{\perp}^j),
\eea
where the full symmetry index (CFT bulk index) is $a=1, 2, \cdots, p$, and the index for the broken directions (associated to the broken generators $h_{\perp}$) is $j=1, 2, \cdots, d-p$.
The $\tau\in\mathbb{R}^p$ are the coordinates along the defect, and $x_{\perp}\in\mathbb{R}^{d-p}$ are the coordinates transverse to the defect.
The defect operators are at $x_{\perp}=0$.
The bare defect operator $W$ is simply the operator that implements the presence of the defect in the path integral before any defect operators are inserted, which is analogous to inserting
\bea
W(1)=\exp\big(-S_{\mathrm{defect}}(\phi)\big)
\eea
on $\mathbb{R}^p$.
\\

\noindent
Starting from the bare defect insertion $W(1)$, one can deform it by turning on sources of the operator $\hat{t}(\tau)$, $w^j(\tau)$,
\bea
W\bigg(e^{\int d^p\tau\ w_j(\tau)\hat{t}_j(\tau)}\bigg),
\eea
which implies the compact notation
\bea
W(\cdots)=\sum_{n=0}^{\infty}\frac{1}{n!}\int d\tau_1d\tau_2\cdots d\tau_n\ w_{i_1}(\tau_1)w_{i_2}(\tau_2)\cdots w_{i_n}(\tau_n)W(1)\hat{t}_{i_1}\hat{t}_{i_2}\cdots\hat{t}_{i_n}.
\eea
\\

\noindent
The presence of the local defect breaks the symmetry group from $G$ to $H$, which means that only a subgroup $H$ of the CFT bulk symmetry survives.
The global symmetry is explicitly broken, and locality is violated along the defect directions.
The current conservation equation must be changed only at $x_{\perp}=0$, which forces a delta function structure
\bea
\partial_{\mu}J^{\mu}_a(x)=\delta^{(d-p)}(x_{\perp})P^j_a\hat{t}_j(\tau),
\eea
where $P^j_a$ is a projector extracting the part of the generator $a$ lying in the boken direction $j$.
This is a universal identity for internal-symmetry breaking by a local defect.
The local defect has degrees of freedom that encode its response to the broken symmetry.
The $\hat{t}_j(\tau)$ is the response that represents the infinitesimal action of the broken internal symmetry from the local defect operators.
Hence, the name tilt operator.

\subsection{Generating Function}
\noindent
The generating functional is \cite{Belton:2025ief}
\bea
Z[r, w]=\int{\cal D}\phi\ e^{-S}\exp\bigg(\int d^dx\ r_{\alpha}(x){\cal O}_{\alpha}(x)\bigg)W\bigg(e^{\int d^p\tau\ w_j(\tau)\hat{t}_j(\tau)}\bigg),
\eea
where $S$ is the CFT bulk action, and ${\cal O}_{\alpha}$ are the CFT bulk correlators.
We can use functional derivatives to consider the most general correlators that do not involve global defect operators.
\\

\noindent
A group element $g\in G$ acts on the sources by a map
\bea
(r, w)\mapsto (L_gr, L_g w).
\eea
For an infinitesimal transformation $g=\exp(\lambda)$ with $\lambda$ in the Lie algebra, we define:
\bea
L_{e^{\lambda}}r=r+\rho(\lambda, r)+{\cal O}(\lambda^2); \
L_{e^{\lambda}}w=w+l(\lambda, w)+{\cal O}(\lambda^2),
\eea
where $\rho(\lambda, r)$ and $l(\lambda, w)$ are the infinitesimal variations of the CFT bulk and local defect couplings.
The local defect variation $l(\lambda, w)$ is expanded as a formal power series in $w$,
\bea
l(\lambda, w)=\sum_{k=0}^{\infty}\frac{1}{k!}l_k(\lambda; w, w, \cdots, w),
\eea
where $l_k(\lambda; w, w, \cdots, w)$ is a multi-linear functional of $k$ copies of $w$.
The $l_0(\lambda)$ is independent of $w$, which is just a vector in the broken directions, and the $l_1(\lambda; w)$ is linear in $w$, etc.
\\

\noindent
We first introduce the shorthand notation:
\bea
{\cal O}(r)\equiv\int d^dx\ r_{\alpha}(x){\cal O}_{\alpha}(x); \
\hat{t}(w)\equiv\int d^p\tau\ w_j(\tau)\hat{t}_j(\tau).
\eea
The logarithm of the generating functional about the connected correlators is
\bea
\ln Z[r, w]=\sum_{m, n\ge 0}\frac{1}{m!n!}\big\langle \big({\cal O}(r)\big)^m\big(\hat{t}(w)\big)^n\big\rangle_c,
\eea
where the coefficient $1/(m!n!)$ in this exapnsion is precisely the connected correlator with $m$ insertions of ${\cal O}(r)$ and $n$ insertions of $\hat{t}(w)$.
Under the infinitesimal transformation, the logarithm of the partition function becomes \cite{Belton:2025ief}
\bea
\ln Z[L_gr, L_g w]=\ln Z[r, w]+ A[\lambda, w],
\eea
where $A[\lambda, w]$ is a defect-local anomaly functional, which can be expanded as
\bea
A[\lambda, w]=\sum_{n=1}^{\infty}\frac{1}{n!}A_n[\lambda; w, w, \cdots, w],
\eea
where each $A_n[\lambda; w_1, w_2, \cdots, w_n]$ is multilinear and symmetric in the $w_j$'s.
\\

\noindent
If we now set $w=0$, it corresponds to turning off all explicit symmetry-breaking local defect couplings
\bea
\ln Z[L_gr, 0]=\ln Z[r, 0]+ A_0[\lambda],
\eea
where
\bea
A_0[\lambda]=A[\lambda, w=0].
\eea
In this situation, by assumption, the theory (bulk + local defect with $w=0$) has an exact global symmetry $G$.
This means that the generating functional is invariant, which implies \cite{Belton:2025ief}
\bea
A_0[\lambda, 0]=0
\eea
Hence, the $A_0$ is simply the anomaly evaluated at zero defect source, and since the theory with $w=0$ is genuinely symmetric (no defect couplings, no anomaly), that anomaly must vanish.
The $A_1$ is linear in $w$, which can be eliminated by a local redefinition of the local defect couplings (a local counterterm), so only $A_{n\ge 2}$ contains non-trivial physical anomaly data \cite{Belton:2025ief}.

\subsection{Integral Identity}
\noindent
We expand the logarithm of the partition function up to the first order in $\lambda$,
\bea
\delta_{\lambda}\ln Z=A[\lambda, w],
\eea
where
\bea
\delta_{\lambda}\ln Z=\ln \big(Z[r+\rho(\lambda, r), w+l(\lambda, w)]\big)-\ln \big(Z[r, w]\big),
\eea
and it is linear in $\lambda$.
We can now plug in the expansion
\bea
\ln Z[r, w]=\sum_{m, n\ge 0}\frac{1}{m!n!}\big\langle \big({\cal O}(r)\big)^m\big(\hat{t}(w)\big)^n\big\rangle_c,
\eea
for $\ln Z$ and keep only the linear terms in $\lambda$.
To consider the variation of the bulk resource \cite{Belton:2025ief}
\bea
r\rightarrow r+\rho(\lambda, r)
\eea
and up to the first-order in $\lambda$, we obtain \cite{Belton:2025ief}
\bea
\ln Z[r+\rho(\lambda, r), w]=\sum_{m, n\ge 0}\bigg\langle\bigg({\cal O}(r)+{\cal O}\big(\rho(\lambda, r)\big)\bigg)^m\big(\hat{t}(w)\big)^n\bigg\rangle_c.
\eea
Hence, the bulk variation of the logarithm of the partition function is \cite{Belton:2025ief}
\bea
\delta_{\lambda}^{\mathrm{bulk}}\ln Z
=
\sum_{m, n\ge0}\frac{1}{m!n!} \big\langle{\cal O}\big(\rho(\lambda, r)\big)\big({\cal O}(r)\big)^{m}\big(\hat{t}(w)\big)^n\big\rangle_c,
\eea
where $\langle\cdots\rangle_c$ is the expectation value of the connected correlators.
We can extract the term \cite{Belton:2025ief}
\bea
\frac{1}{n!}\sum_{j=1}^m\big\langle{\cal O}(r_1){\cal O}(r_2)\cdots{\cal O}\big(\rho(\lambda, r_j)\big)\cdots {\cal O}(r_m)\big(\hat{t}(w)\big)^n\big\rangle_c.
\eea
\\

\noindent
We now vary the local defect source \cite{Belton:2025ief}
\bea
w\rightarrow w+l(\lambda, w)
\eea
up to the first-order in $\lambda$ to show \cite{Belton:2025ief}
\bea
\ln Z[r, w+l(\lambda, w)]=\sum_{m, n\ge 0}\frac{1}{m!n!}\bigg\langle\big({\cal O}(r)\big)^m\bigg(\hat{t}(w)+\hat{t}\big(l(\lambda, w)\big)\bigg)^n\bigg\rangle_c.
\eea
The local defect variation of the logarithm of the partition function is \cite{Belton:2025ief}
\bea
\delta_{\lambda}^{\mathrm{defect}}\ln Z
=\sum_{m, n\ge 0}\frac{1}{m! n!}\sum_{k\ge 0}\frac{1}{k!}\big\langle\big({\cal O}(r))^m\hat{t}\big(l_k(\lambda; w, w, \cdots, w)\big)\big(\hat{t}(w)\big)^{n}\big\rangle_c.
\eea
We extract the terms with a fixed-order in $n$ for $w$ \cite{Belton:2025ief},
\bea
\frac{1}{n!}\sum_{k=0}^n\frac{n!}{k!(n-k)!}\big\langle{\cal O}(r_1){\cal O}(r_2)\cdots{\cal O}(r_m)\hat{t}\big(l_k(\lambda; w, w, \cdots, w)\big)\big(\hat{t}(w)\big)^{n-k}\big\rangle_c.
\eea
The coefficient is given by the $k$ choice in the $l_k(\lambda; w, w, \cdots, w)$ and also the remaining $n-k$ choices in $\hat{t}(w)$ \cite{Belton:2025ief}.
We combine the CFT bulk and local defect variation of the sources and the defect anomaly, and we then obtain the integral identities for the connected correlators \cite{Belton:2025ief}
\bea
&&
\sum_{j=1}^m\big\langle{\cal O}(r_1){\cal O}(r_2)\cdots{\cal O}\big(\rho(\lambda, r_j)\big)\cdots {\cal O}(r_m)\big(\hat{t}(w)\big)^n\big\rangle_c
\nn\\
&&
+
\sum_{k=0}^n\frac{n!}{k!(n-k)!}\big\langle{\cal O}(r_1){\cal O}(r_2)\cdots{\cal O}(r_m)\hat{t}\big(l_k(\lambda; w, w, \cdots, w)\big)\big(\hat{t}(w)\big)^{n-k}\big\rangle_c
\nn\\
&=&A_n[\lambda; w, w, \cdots, w].
\eea
\\

\noindent
Let us illustrate some examples.
For $n=0$, the sum over $k$ has only one term, $k=0$ \cite{Belton:2025ief},
\bea
\langle\hat{t}\big(l_0(\lambda)\big)\rangle=A_0[\lambda].
\eea
Because $l_0(\lambda)$ is the $w$-independet part of the defect source variation, we get \cite{Belton:2025ief}
\bea
\hat{t}\big(l_0(\lambda)\big)=\hat{t}(\lambda).
\eea
Hence, we get \cite{Belton:2025ief}
\bea
\langle \hat{t}(\lambda)\rangle_c=0.
\eea
We now do the variation for $\lambda_j$ and then set $\lambda=0$ to show \cite{Belton:2025ief}
\bea
\frac{\delta}{\delta\lambda_j}\langle\hat{t}(\lambda)\rangle_c\bigg|_{\lambda=0}
=\int d^p\tau\ \langle \hat{t}_j(\tau)\rangle_c=0.
\label{id1}
\eea
When considering the conformal defect group, the one-point tilt function vanishes
\bea
\langle t_j(\tau)\rangle=0,
\eea
which automatically satisfy the integral identity \cite{Belton:2025ief}.
Now, we go to $n=1$ for the defect-only identity \cite{Belton:2025ief},
\bea
\big\langle\hat{t}(\lambda)\hat{t}(w)\big\rangle_c
+\big\langle \hat{t}\big(l_1(\lambda; w)\big)\big\rangle_c
=0.
\eea
We continue to consider the conformal defect group case, and the defect one-point function vanishes, and then it shows  \cite{Belton:2025ief}
\bea
\big\langle\hat{t}(\lambda)\hat{t}(w)\big\rangle_c
=0.
\eea
We then take functional derivatives for the sources and then set them to zero \cite{Belton:2025ief},
\bea
\frac{\delta}{\delta\omega_k(\tau_2)}\frac{\delta}{\delta\lambda_j}\langle\hat{t}(\lambda)\hat{t}(w)\rangle_c\bigg|_{\lambda=w=0}
=\int d^p\tau_1\ \langle\hat{t}_j(\tau_1)\hat{t}_k(\tau_2)\rangle_c=0.
\eea
Now, we get an integral constraint to the two-point tilt function \cite{Belton:2025ief}
\bea
\int d^p\tau_1\ \langle\hat{t}_j(\tau_1)\hat{t}_k(\tau_2)\rangle_c=0.
\label{id2}
\eea
\\

\noindent
Let us now consider the integral identity for $(m, n)=(1, 0)$ \cite{Belton:2025ief},
\bea
\big\langle{\cal O}\big(\rho(\lambda, r)\big)\big\rangle_c
+\langle{\cal O}(r)\hat{t}(\lambda)\rangle_c
=0.
\eea
The infinitesimal action on the bulk source is \cite{Belton:2025ief}
\bea
\rho(\lambda, r)_{\beta}(x)=\lambda_jr_{\alpha}(x)(T_j)_{\beta\alpha},
\eea
where $(T_j)_{\beta\alpha}$ are the representation matrices of the broken generators.
We take the functional derivatives \cite{Belton:2025ief}:
\bea
\frac{\delta}{\delta\lambda_j}\frac{\delta}{\delta r_{\alpha}(x_0)}\big\langle{\cal O}\big(\rho(\lambda, r)\big)\big\rangle_c\bigg|_{\lambda=r=0}
&=&(T_j)_{\beta\alpha}\langle{\cal O}_{\beta}(x_0)\rangle_c; \
\nn\\
\frac{\delta}{\delta\lambda_j}\frac{\delta}{\delta r_{\alpha}(x_0)}\langle{\cal O}(r)\hat{t}(\lambda)\rangle_c\bigg|_{\lambda=r=0}
&=&\int d^p\tau_1\ \langle{\cal O}_{\alpha}(x_0)\hat{t}_j(\tau_1)\rangle_c.
\eea
Combining the results that we get shows that \cite{Belton:2025ief}
\bea
(T_j)_{\beta\alpha}\langle{\cal O}_{\beta}(x_0)\rangle_c+\int d^p\tau_1\ \langle{\cal O}_{\alpha}(x_0)\hat{t}_j(\tau_1)\rangle_c=0.
\label{id3}
\eea
The result implies that the generator acting on the one-point function plus integrated bulk-tilt two-point function is zero.
Hence, our examples demonstrate the constraints on the connected correlators and how the connected correlator can be determined from the lower-point connected correlator \cite{Belton:2025ief}.

\subsection{Conformal Defect Group}
\noindent
To apply the conformal symmetry to determine the low-point correlators, we consider the infinite planar defect.
The full conformal group in $d$ dimensions is SO($d+1$, $1$).
A flat $p$-dimensional local defect preserves a conformal group along the defect, SO($p+1$, 1), which includes
\begin{itemize}
\item{translations along the defect
\bea
\tau^a\rightarrow\tau^a+c^a;
\eea
}
\item{rotations within the defect SO($p$) and rotations in directions normal to the defect, SO($d-p$);}
\item{dilatations: $(\tau, x_{\perp})\rightarrow(\lambda\tau, \lambda x_{\perp})$.}
\item{special conformal transformations tangent to the defect.}
\end{itemize}
Hence, the preserved symmetry group is SO($p+1, 1$)$\times$SO($d-p$), which is the standard "defect conformal group".
The title operator $\hat{t}_j$ arises because the defect breaks an internal symmetry $G\rightarrow H$.
For the broken generators $j\in h_{\perp}$, the current conservation equation becomes the defect-localized Ward identity
\bea
\partial_{\mu}J^{\mu}_j(x)=\delta^{(d-p)}(x_{\perp})\hat{t}_j(\tau).
\eea
Hence, the tilt operator is a local operator with scaling dimension $\Delta_{\hat{t}}=p>0$ that lives on the defect.
It transforms as a vector in the broken symmetry space $h_{\perp}$.
\\

\noindent
Since we have the conformal group for each CFT bulk and defect space, respectively, we can introduce the same adjoint relations to the generators for the CFT bulk and defect direction indices as our previous analysis (whole conformal group in CFT$_d$), respectively.
To compute higher-point correlators with a proper inner product, we can introduce the global defect only via an invariant state under a global defect operator.
Suppose a state is invariant under a global defect operator.
In that case, we can take its complex conjugate to see how the global defect must be introduced.
Because the vacuum state is not an invariant state under a global defect operator, the CFT bulk or defect operators need to act on the vacuum state via a linear combination to form an invariant state under a global defect operator.
Hence, only the operator acting on the bra state is affected by a global defect operator.
Another reason for twisting the inner product in CFT is that the dS bulk operator can be translated to the CFT operators through the HKLL bulk reconstruction procedure \cite{Brown:1986nw,Cotler:2018zff,Huang:2019nfm,Huang:2020tjl,Bhowmick:2019nso}.
It should imply that the computation of CFT correlators requires the introduction of the same global defect to twist the inner product.
Because we twist the inner product from a global defect, the symmetry needs to be enlarged.
If we consider the global defect operator as $PT$, we can enlarge the defect conformal group to O($p+1$, $1$)$\times$O($d-p$).
\\

\noindent
When considering the defect conformal group, we have the maximum possible conformal symmetry.
We can use translational symmetry to constrain the expectation value of the one-point tilt operator to be a constant.
We then use the scaling symmetry to conclude that the constant must be zero, thereby automatically satisfying the integral constraint \eqref{id1}.
Hence, the defect still explicitly breaks the symmetry, rather than breaking it spontaneously.
\\

\noindent
For the two-point tilt operators, we can first consider translational and rotational symmetries along the defect directions to show that the two-point Green's function of the tile operators depends only on the distance in defect coordinates.
We then apply the scaling to the two-point tilt operators, which implies
\bea
\langle\hat{t}_j(\tau_1)\hat{t}_k(\tau_2)\rangle_c=\frac{C_{jk}}{|\tau_{12}|^{2p}},
\eea
where
\bea
\tau_{12}\equiv\tau_1-\tau_2.
\eea
Because the tile operators transform as a vector under the internal representation, and the vacuum state is invariant under the internal symmetry, the only invariant tensor is proportional to the identity matrix
\bea
C_{jk}=C_t\delta_{jk}.
\eea
Hence, the two-point tilt function also satisfies the integral constraint \eqref{id2}.
Let us specifically consider the global defect operator $PT$.
Only the parity operator acts on the defect directions. The expectational value of the two-point tilt operators does not vanish only when two operators have the same parity.
The same analysis applies to other one- and two-point correlators, and the integral constraints help connect higher-point connected correlators, such as Eq. \eqref{id3}.
Hence, the introduction of global defect operators enlarges the symmetry group and imposes additional constraints on the CFT correlators.
The introduction of the global defect operator, which twists the inner product, simplifies the computation of correlators.

\section{Discussion and Conclusion}
\label{sec:7}
\noindent
In this work, we have investigated the symmetry-based perspective of the dS/CFT correspondence using Wick rotation and analytic continuation.
Beginning with global AdS$_{d+1}$ in Lorentzian signature, we demonstrated how these operations yield the static patch of dS$_{d+1}$ together with the appropriate bulk dS$_{d+1}$ isometry group, from which one can read off the corresponding CFT$_d$ generators.
The resulting adjoint operation for these generators in the dS bulk coordinate representation differs from the familiar AdS/CFT case, leading to an exotic adjoint structure that naturally introduces a global defect operator $PT$.
This operator is essential for reproducing the correct Green’s function of a real scalar field in dS$_{d+1}$.
Motivated by the antipodal symmetry inherent to the sphere, we posited that vacuum degeneracy should be anticipated within the dS/CFT framework. 
Furthermore, we explicitly demonstrated that any CFT$_2$ with a central charge possessing an imaginary component must necessarily exhibit spontaneous $PT$-symmetry breaking of its vacuum state. 
Finally, we scrutinized connected correlators in the presence of both global and local defects using integral identities, elucidating how global defect operators can be seamlessly incorporated into these identities and how the defect conformal group must be appropriately expanded in the dS/CFT correspondence.
\\

\noindent
We applied Wick rotation and analytic continuation only to the metric, CFT generators, and Green’s functions, and we found that this restricted procedure works consistently in arbitrary dimensions.
However, these operations remain subtle in general quantum field theories, particularly in curved spacetime.
When applied directly to the Lagrangian, difficulties arise in the path integral and regularization.
Our examination reveals that Wick rotation and analytic continuation can be executed reliably on numerical quantities—such as Green’s functions post-evaluation—yet cannot be applied directly to bra states. 
While the continuation generates expressions akin to Eq.~\eqref{gf1}, such expressions ought to correspond to expectation values of vacuum states whose conjugate spectra are distinct, thereby indicating that the continuation cannot be uniformly applied to all states.
Instead, the exotic adjoint structure we introduced provides a consistent realization of observables through the Wick rotation and analytical continuation.
This observation is promising for the study of dS$_{d+1}$ quantum gravity from CFT$_d$ data.
Once expectation values of all connected correlators are known, the dS bulk theory is fully determined, even without explicit knowledge of its Lagrangian. 
The central lesson of this work is that analytic continuation from AdS to dS modifies not only correlation functions but also the notion of adjointness and the Hilbert-space metric. 
The resulting $PT$-defect structure provides a possible bridge between analytic continuation approaches, non-unitary dS dualities \cite{Strominger:2001pn}, and higher-spin realizations of dS/CFT \cite{Anninos:2011ui}. 
\\

\noindent
Our analysis revealed the necessity of introducing a $PT$ operator from the two-point function of a real scalar field in a de Sitter background. Since charge conjugation is trivial for real scalars, a more general global defect structure likely requires considering fields with nontrivial spin or internal charges.
The $PT$ operator is consequently anticipated to be exclusive to real scalar fields, and supplementary global defects will necessitate corresponding augmentations of the conformal symmetry group. Furthermore, it is imperative to examine higher-point connected correlators—such as conformal blocks—beyond the realm of the two-point de Sitter bulk Green’s function. 
While the two-point function can be portrayed as the inner product of two bulk local states, this assertion does not hold for higher-point functions; thus, comprehending their structure is vital for ascertaining how the inner product ought to be twisted in the de Sitter/conformal field theory correspondence.
Such higher-point data also encode information about interactions in the dS bulk theory.
For a well-defined theory, the global defect should act as a metric operator \cite{Mostafazadeh:2001jk}.
However, we focused on $PT$ symmetry because the free theory respects it; interacting theories may require different choices.
Exploring these possibilities provides a compelling direction for formulating a consistent, physically meaningful dS quantum gravity theory from CFT principles.
\\

\noindent
 In the $PT$-broken phase of our non-Hermitian defect CFT, the one-point function of the tilt operator acquires a nonzero value, signaling that the vacuum is no longer invariant under the combined defect conformal symmetry and $PT$ symmetry. 
This nonvanishing one-point tilt function acts as an order parameter that selects a specific $PT$-broken vacuum from a degenerate manifold, encoding how the defect sources bulk fields and breaks scale invariance along the defect worldvolume. 
Because the AdS and dS situations differ substantially in both symmetry and analytic structure, this disorder parameter provides a natural diagnostic for distinguishing the two phases.  
Notably, although the background expectation value reflects spontaneous $PT$-symmetry breaking, the fluctuating part of the defect operator continues to transform covariantly under the unbroken subgroup of the defect conformal group. 
Consequently, higher-point connected correlators—such as two-point functions of defect fluctuations—remain constrained by standard defect conformal symmetry, even when the whole theory resides in a symmetry-broken phase. 
Thus, while the nonzero one-point function enriches defect physics by introducing a symmetry-breaking background, the conformally invariant structure of fluctuations persists. 
The $PT$-breaking studied here represents the simplest example of a nonvanishing one-point tilt function beyond the conventional defect conformal group. 
It would be compelling to apply the integral identity of Ref. \cite{Belton:2025ief} without assuming the vanishing one-point tilt function to analyze higher-point functions in this expanded framework.
\\

\noindent
 In this work, we have realized the dS/CFT correspondence for heavy bulk particles whose conformal dimensions possess imaginary parts. 
In contrast, light particles have real-valued conformal dimensions $\Delta$ and $d-\Delta$.  
This structure is precisely what appears in the study of light fields in cosmology, where such mixed-dimension behavior plays a central role in inflationary correlators. 
It would therefore be exciting to explore the interplay between bulk reconstruction and non-Hermitian defect structures in the light-particle regime of the dS/CFT correspondence.

\section*{Acknowledgments}
\noindent 
We thank Ziwen Kong, Tadashi Takayanagi, and Gerard Watts for their helpful discussion.
XH acknowledges the NSFC Grants (Grants No. 12247103 and No. 12475072).
CTM thanks Nan-Peng Ma for his encouragement.

%\noindent
%The author acknowledges the YST Program of the APCTP; 
%Post-Doctoral International Exchange Program (Grant No. YJ20180087); 
%China Postdoctoral Science Foundation, Postdoctoral General Funding: Second Class (Grant No. 2019M652926); 
%Foreign Young Talents Program (Grant No. QN20200230017); 
%Science and Technology Program of Guangzhou (Grant No. 2019050001).

%\appendix

\end{document}